\documentclass[conference]{IEEEtran}
\IEEEoverridecommandlockouts
\usepackage{cite}
\usepackage{amsmath,amssymb,amsfonts}
\usepackage{algorithmic}
\usepackage{graphicx}
\usepackage{textcomp}
\usepackage{xcolor}
\usepackage{multirow}
\usepackage{tabularx}
\def\BibTeX{{\rm B\kern-.05em{\sc i\kern-.025em b}\kern-.08em
    T\kern-.1667em\lower.7ex\hbox{E}\kern-.125emX}}
\begin{document}

\title{Stochastic Communication Avoidance for Recommendation Systems\\}
\author{
\IEEEauthorblockN{Lutfi Eren Erdogan\IEEEauthorrefmark{1}\textsuperscript{1}, Vijay Anand Raghava Kanakagiri\IEEEauthorrefmark{1}\textsuperscript{2}, Kurt Keutzer\textsuperscript{1}, Zhen Dong\textsuperscript{1}}
\IEEEauthorblockA{lerdogan@berkeley.edu, vijay7anand@tamu.edu, keutzer@berkeley.edu, zhendong@berkeley.edu}
}

\maketitle
\begingroup
\renewcommand\thefootnote{}
\footnotetext{-----------------}
\footnotetext{*Equal contribution, \textsuperscript{1}University of California Berkeley, \textsuperscript{2}Texas A\&M University}
\endgroup
\begin{abstract}
\textit{
One of the major bottlenecks for efficient deployment of neural network based recommendation systems is the memory footprint of their embedding tables.
Although many neural network based recommendation systems could benefit from the faster on-chip memory %
access and increased computational power of hardware accelerators, the large embedding tables in these models often cannot fit on the constrained memory of accelerators.
Despite the pervasiveness of these models, 
prior methods in memory optimization and parallelism fail to address the memory and communication costs of large embedding tables on accelerators.
As a result, the majority of models are trained on CPUs, while current implementations of accelerators are hindered by issues such as bottlenecks in inter-device communication and main memory lookups.
In this paper, we propose a theoretical framework that analyses the communication costs of arbitrary distributed systems that use lookup tables.
We use this framework to propose algorithms that maximize throughput subject to memory, computation, and communication constraints.
Furthermore, we demonstrate that our method achieves strong theoretical performance across dataset distributions and memory constraints,
applicable to a wide range of use cases from mobile federated learning to warehouse-scale computation.
We implement our framework and algorithms in PyTorch and
achieve up to $6\times$ increases in training throughput on GPU systems over baselines, on the Criteo Terabytes dataset.
}
\end{abstract}

\section{Introduction}

A significant portion of machine learning research has advanced due to better memory and computational speeds of accelerators, alongside faster interconnects and more efficient parallelization in large systems. However, accelerators often have limited memory compared to CPUs, rendering many memory-intensive algorithms infeasible for deployment. One approach to mitigate this issue is to increase memory, but this cannot keep up with the rapid growth of machine learning models. An alternative is to develop new parallelization strategies that balance memory usage and communication, as explored in strategies like those in~\cite{fan2021dapple, jia2019beyond, zhao2023pytorch}. Another optimization strategy involves quantization~\cite{gholami2022survey, dong2019hawq, liu2023noisyquant}, which aims to minimize the memory footprint and computational requirements without significantly impacting accuracy. Embedding tables, which map sparse categorical features to dense vectors~\cite{cheng2016wide, zhao2022analysis}, are often prime targets for quantization due to their large sizes. By quantizing the embeddings to lower bit-width representations in DLRMs~\cite{naumov2019deep}, such as 4-bit~\cite{guan2019post, zhou2024dqrm}, or performing tensor train decomposition~\cite{yin2021tt}, memory usage can be significantly reduced, making it more feasible to train and inference. These methods primarily focus on reducing the embedding size~\cite{li2024embedding}, while another solution is to modify the training process to save memory. Examples of this approach include reversible networks~\cite{mangalam2022reversible}, which change the model’s structure, and techniques like Checkmate~\cite{jain2020checkmate}, which alter the model’s execution pattern by adding additional operations to backpropagation to decrease the number of intermediate values that need to be stored in memory.

Recent algorithms have significantly increased the scale of NLP and CV models by reducing the memory demands per GPU~\cite{shang2023pb,li2023qft, shoeybi2019megatron}, allowing the use of accelerators for extremely large models. However, these methods struggle with models that have large embedding tables, which are not easily managed by pipeline parallelism and remain large in parameter count. Data parallelism also falls short as it is better suited for compute-heavy tasks rather than memory-intensive embedding operations. Furthermore, techniques like recomputation or checkpointing do not suit embeddings well, as their high memory cost does not justify the modest savings from managing intermediate activations.

This forces the use of model parallelism, but oftentimes the number of accelerators required to fit the large embedding tables is too great to make model parallelism a financially viable solution.  

These embeddings can often be terabytes large, but as observed in practice, they are not accessed uniformly at random.
In real datasets, the access pattern of these embeddings varies, generally with a small portion of embeddings being accessed far more frequently than others\cite{agarwal2023bagpipe}.
Existing methods~\cite{zhao2020distributed}
have explored the usage of distributed communication to decrease the
communication cost, but the theoretical bounds of communication
efficiency has not been analyzed in previous work. In addition, there has been no existing work exploring how various methods
of distributing embeddings across GPUs and CPUs impact the system performance.

To summarize, our contributions are as follows:
\begin{enumerate}
    \item
    We develop a simple framework to calculate the expected communication cost
    under a training regime. %
    And we expand this framework to address considerations such as determining the optimal levels of communication and caching, as well as methods for adjusting them accordingly.
    
    \item
    We use the above framework to obtain communication strategies that minimize expected communication costs without caching. We demonstrate that these methods also decrease main memory I/O proportionally to the decrease in communication.
    
    \item 
    We demonstrate how assumptions about ML training motivate different strategies
    for caching through empirical analysis with our framework.
    
    \item 
    We extensively test our algorithms on a variety of datasets and models.
    Furthermore, we also test various theoretical distributions and observe that our
    algorithms can generalize well, with performance gains on a wide range of potential 
    synthetic datasets.
\end{enumerate}

\section{Methodology}

\begin{figure}[htbp]
\centerline{\includegraphics[width=0.5\textwidth]{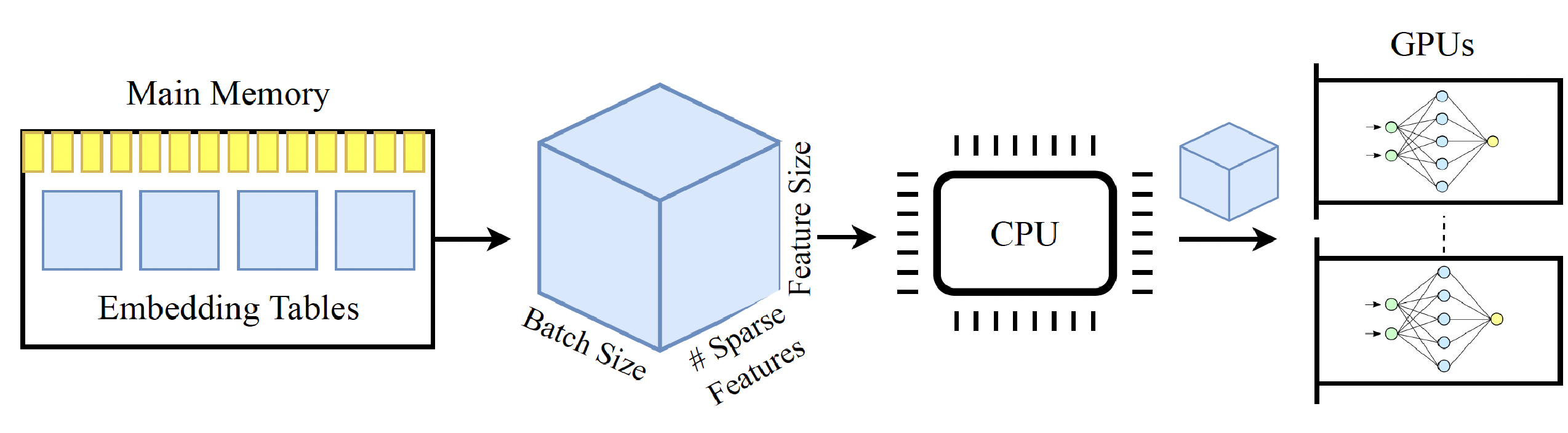}}
\caption{Existing training paradigm without coalescing.}
\label{fig:baseline}
\end{figure}

As shown in Fig.~\ref{fig:baseline}, the existing training paradigm for the above model is as follows
\begin{enumerate}
    \item A batch of training examples is retrieved
    \item For every training example, the embeddings are retrieved from embedding tables in the main memory
    \item The batch of embeddings is sent to the GPU
\end{enumerate}

In the following chapters, we will analyze the communication cost of the embedding layers of the recommendation systems and propose algorithms that maximize throughput subject to memory, computation, and communication constraints.  

\subsection{Coalescing on device}

To start, we wish to estimate how much overlap exists per batch of the dataset. To do so, we want to find the expected number of unique elements in the list and the expected total communication given a batch size $b$. If we transmit the batch of $b$ training examples, the communication cost per feature is
$b\times d$, where d is the number of lookups per sample.
We call this "communicating the batch", as we send over the entire batch.

\begin{figure}[htbp]
\centerline{\includegraphics[width=0.5\textwidth]{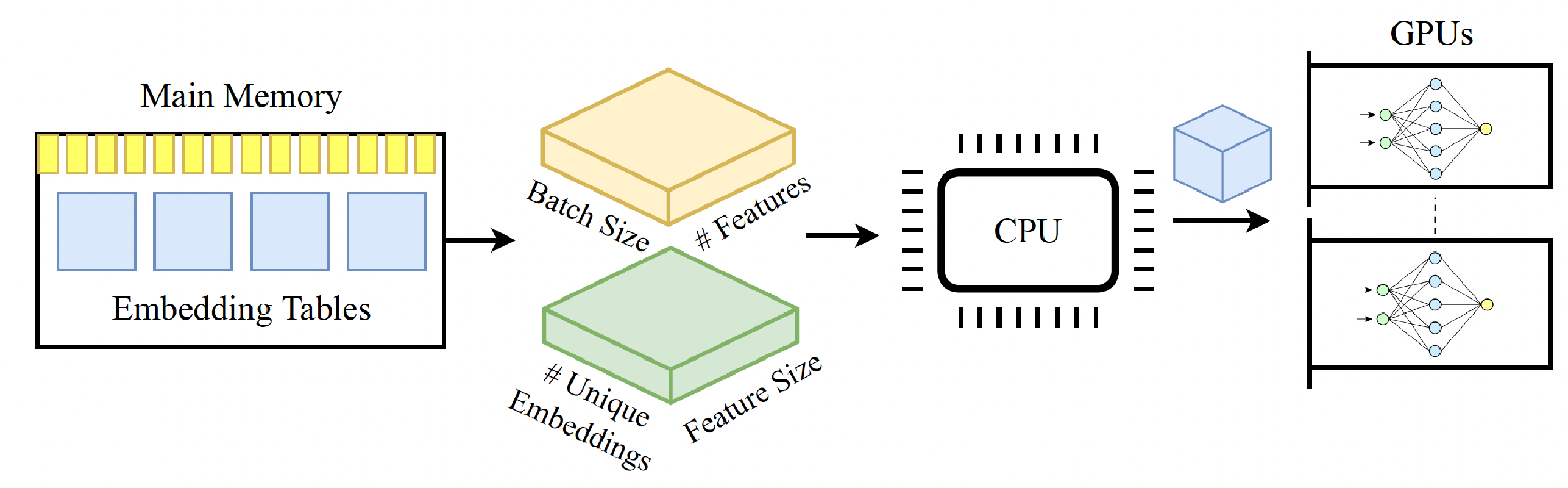}}
\caption{Training paradigm with coalescing.}
\label{fig:coaslecing}
\end{figure}

As depicted in Fig.~\ref{fig:coaslecing} we can alternatively coalesce the embeddings into a list of unique embeddings in the batch as well as their indices. 
For embedding e, let $P(e)$ be the probability that e refers to a categorical feature in a training example chosen uniformly at random from the dataset.
The likelihood that $e$ is transmitted within our batch is computed by taking the complement of the probability that no embedding for that feature in our batch is e:
\begin{equation}
1 - (1-P(e))^b
\label{eq:1}
\end{equation}
Using (\ref{eq:1}), if $E$ is the set of possible embeddings for this feature, we can then find our expected embedding communication cost as:
\begin{equation}
    \sum_{e \in E} 1 - (1-P(e))^b
\label{eq:2}
\end{equation}

In addition to transmitting the unique embeddings as described in (\ref{eq:2}), we also have to transmit the indices of each embedding. 
This means that we have a constant $b$ cost. As a result, our net cost is
\begin{equation}
    b + \sum_{e \in E} 1 - (1-P(e))^b
\label{eq:3}
\end{equation}

We call this method "coalescing" because coalesces, or combines distinct, embeddings for each feature.

\subsection{Caching on device}

\begin{figure}[htbp]
\centerline{\includegraphics[width=0.5\textwidth]{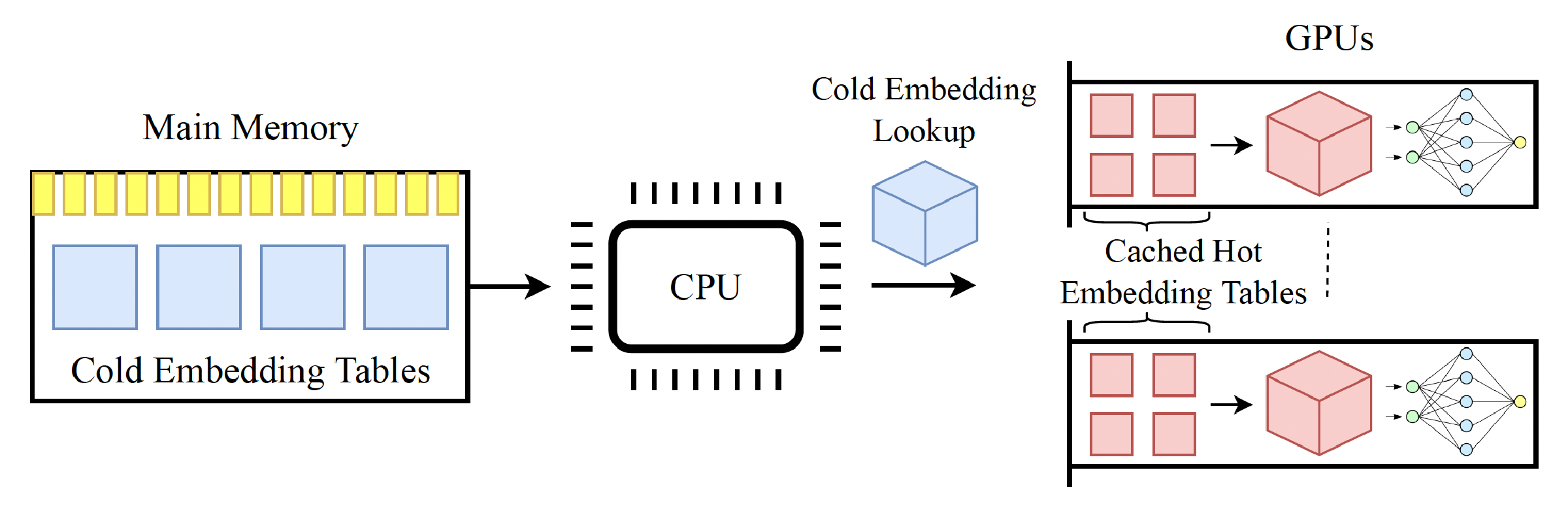}}
\caption{Training Framework with Caching Mechanism.}
\label{fig:caching}
\end{figure}

We can also use the GPU as a cache for our embeddings as described in Fig.~\ref{fig:caching}, we examine several ways of determining what embeddings to store on GPU.\\

The first method: minimize communication bandwidth per epoch. For this method, we assume that the batch size is sufficiently large on GPU to saturate communication during training, and as a result, the time it takes to execute computations using larger batch sizes is proportional to the batch size itself.
As a result, the expected time spent on computation per epoch is constant, and the primary source of change in the latency is how much time is spent on communication per epoch.

We shall let the number of samples in the dataset be $Q$, batch size $b$, and the lookups per sample be $d$.
Our expected communication cost is equal to the expected number of batches times the expected communication per batch. Without caching or overlap, this is equal to:
\begin{equation}
    \frac{Q}{b} \times b \times d
    = Q \times d
\label{eq:4}
\end{equation}

If we exploit the overlap between lookups, the communication cost changes to
\begin{equation}
    Q + \frac{Q}{b}(\sum_{e \in E} 1 - (1-P(e))^b) \times d
\label{eq:5}
\end{equation}
The first term in (\ref{eq:5}) represents the cost of sending indices, while the second term is the cost of sending the embeddings. 
Generally speaking, the communication cost without caching will decrease as batch size increases, but the memory requirements will also increase.
The tradeoff between these two values depends on the distribution of embeddings.

If we utilize the remaining memory to cache embeddings, we can remove
the communication cost of embeddings that are cached on device.
Let the set of embeddings cached on the device be $C$
\begin{equation}
    Q + \frac{Q}{b}(\sum_{e \in E / C} 1 - (1-P(e))^b) \times d
\label{eq:6}
\end{equation}
However, due to the fact that we have limited memory, there is a direct trade-off between the batch size and the number of cached embeddings.

As a result, if we model the memory usage of the cached embeddings, we can theoretically calculate the largest potential batch size on device.
While in practice compilers do not allocate memory with this theoretical efficiency, it helps illustrate the mathematics of the tradeoffs.
Define the total number of parameters that can fit on the device to be $M$, the parameters used for running the model per sample to be $a$, i.e. space taken up by both model's weights, biases and also space required for intermediate activations of a single sample.
Then, given that $|C|$ embeddings are cached, and each take up $d$ parameters on device, as before,
the maximum batch size possible is
\begin{equation}
    b = \frac{M - |C|d}{a}
\label{eq:7}
\end{equation}
This introduces a tradeoff between the amount of communication saved by overlap and the amount of communication saved by caching, because the higher our batch size is the more overlap will occur but the less memory will be available for caching.

In order to understand the efficacy of this tradeoff, we need to examine the relative change in communication from caching one additional embedding, $e'$.

Let C be the current set of cached embeddings and $b$ be the maximum batch size with $C$ cached. 

Let $C' = C \cup e'$, $b'$ be be the maximum batch size with $C'$ cached, 

If we store $e'$, our expected decrease in communication cost is $1 - (1-P(e))^b$, which is the likelihood $e'$ is in a batch.

However, using (\ref{eq:7}), our expected batch size decreases by ${b-b'} = \frac{d}{a}$, or the size of one embedding divided by the size of activation.
This both decreases communication and decreases potential overlap, using (\ref{eq:6}) and (\ref{eq:7}), we can mathematically say that

\begin{equation}
\Delta communication/epoch = 
\text{commn}_{1} - \text{commn}_{2}
\label{eq:8}
\end{equation}
where,
\begin{equation}
\text{commn}_{1} = \frac{\sum_{e \in E/C'} (1-(1-P(e))^{b'}) \times Q}{b'}
\label{eq:9}
\end{equation}

\begin{equation}
\text{commn}_{2} = \frac{\sum_{e \in E/C} (1-(1-P(e))^b) \times Q}{b}.
\label{eq:10}
\end{equation}

For the communication to decrease, we need (\ref{eq:8}) to be negative.
We can separate out the term for $e'$ to get that
\begin{equation}
(1-(1-P(e'))^b) \geq
     t1
\label{eq:11}
\end{equation}
where
\begin{equation}
t1 =\sum_{e \in E/C'} \frac{(b(1-(1-P(e))^{b'})-b'(1-(1-P(e))^b))}{b'}
\label{eq:12}
\end{equation}

When $M >> a > d$, or when available memory is significantly larger than the memory needed for running model on a one sample, and the memory needed for running a model on a single sample is larger than the memory needed for an embedding, caching generally improves performance in use cases where the dataset is randomized, such as offline training. Because this equation minimizes communication bandwidth, it only serves as an accurate model of communication when communication is dominated by bandwidth.

In practice, this means the caching generally decreases the cost of communication relative to relying on coalescing for large batch size training. 

In addition to using this to understand the tradeoff between the batch and the number of cached embeddings, it is possible to find the theoretically optimal number of embeddings to minimize communication bandwidth in $O(log(|E|))$ steps using binary search. However, the embedding memory and activation memory need to be measured empirically, as existing machine-learning platforms often allocate more memory than necessary to improve performance.

Furthermore, (\ref{eq:13}) is also a good proxy for the expected number of accesses to main memory for embeddings.  This is because cache performance for these embeddings are poor, and as a result most embeddings communicated over the channel are looked up in main memory.

\begin{equation}
(1-P(e'))^{(b-b')} \geq
     \frac{d}{M-|C'|d}\sum_{e \in E/C'}
\label{eq:13}
\end{equation}

If the batch size is too small to saturate computation after communication, the size of the cached set can be determined by measuring the runtime of the model on various splits of the data. This is much more time-intensive, but can potentially lead to more accurate results.

When analyzed with the different dataset distributions, our method displays strong theoretical performance. We use three distributions, Zipf distribution ($P\left ( x \right ) \sim 1/x = e^{-log\left ( x \right )}$), exponential distribution ($P\left ( x \right ) \sim 1/x = e^{x}$), and half normal distribution ($P\left ( x \right ) \sim 1/x = e^{-x^{2}}$), to scale to $5\times$ number of embeddings and $5\times$ batch size. Half-normal distribution is of particular importance since the Criteo Terabyte Dataset is most similar to that distribution. With our method, the total communication cost increases by $<1.5\times$ for the exponential and the normal distributions and by $<2\times$ for Zipf distribution while with prior methods, the total communication cost increases by $5\times$. That is a $3\times$ increase in theoretical performance.

\section{Experiment Settings}

We evaluate our method using DLRM~\cite{naumov2019deep} on the Criteo Terabyte Dataset. Since our method and our theoretical framework is applicable to any arbitrary distributed system that uses lookup tables, we expect the same performance gains if we have used TBSM~\cite{ishkhanov2020time} on the Alibaba UBA dataset.

Both models, for their respective datasets, are usually trained on CPUs. This is due to the limited memory availability for GPU training. 

We compare our method against baseline Deep Learning Recommendation Model (DLRM) implementations that use only CPUs or a combination of CPUs and GPUs without caching. The CPU-only setup processes the entire computation graph, including the DNN's large tensor operations, on the CPU—tasks for which the CPU is not optimized. The CPU-GPU setup assigns the memory-intensive embedding layer to the CPU and the compute-intensive DNN layers to the GPU, resulting in CPU-GPU communication overhead during the backward pass and when handling intermediate results, which extends the training time.

The effectiveness of caching hot embeddings in mini-batches hinges on using only the cached (hot) embeddings and avoiding the cold embeddings stored on the CPU, thus eliminating data shuffling between the CPU and GPUs during the embedding layer. While it's unrealistic to expect all mini-batches to require only hot embeddings, it is feasible to ensure that most do. This is achieved by classifying training samples into "hot" (those that only need hot embeddings) and "normal" (those that require both hot and cold embeddings). We can then create mini-batches exclusively composed of hot samples, and others of normal samples. Given that hot samples generally predominate, this method maximizes the use of cached embeddings. This classification involves constructing a ranked skew table for all embeddings, selecting the top few embeddings for caching, and then categorizing training samples based on their dependency on these cached embeddings.

For all of the experiments, we benchmark the iteration time and the total training time for one epoch.

\section{Experiments}

We first evaluate our results on the Criteo Terabyte dataset. For this method, we trained a DLRM model with the default parameters on two NVIDIA TITAN RTX with 24GB of memory. For our first experiment, we wanted to devise a simple experiment to verify the benefit of overlapping and caching on the iteration time and the total training time for one epoch. We set the batch size to 2048 for all runs, and we cache 256MB of the hot embeddings with our method.

From Table \ref{table:1}, we can see that models with caching and coalescing achieve large performance improvements relative to baselines. We even argue that without caching, the CPU-GPU implementation is worse than the CPU-only implementation. This means that the state-of-the-art implementations cannot utilize the hardware accelerators, leading to scalability problems; while, caching enables us to integrate fast accelerators into deep recommendation systems training by avoiding CPU-GPU communication.

\begin{table}[htbp]
\caption{Comparing total time in seconds and one epoch for the two baselines and our method with coalescing and caching.}
\centering
\begin{tabular}{|c|c|c|c|c|c|}
\hline
Method & \multicolumn{4}{c|}{\textbf{One Iteration (s)}} &One \\
\cline{2-5} 
 & Fwd & Bckwd & Optzn & \textbf{\textit{Total}} & Epoch(m) \\
\hline
CPU-only                & 12.05 & 15.1 & 41.94 & 69.09 & 38.03 \\ 
CPU-GPU baseline        & 20.37 & 19.14 & 54.17 & 93.68 & 46.12 \\ 
SCARS (ours)            & 50.88 & 5.36 & 0.41 & \textbf{56.65} 
&\textbf{27.12} \\
\hline
\multicolumn{6}{l}{Note: Fwd, Bckwd, Optzn means Forward, Backward, Optimization}

\end{tabular}
\label{table:1}
\end{table}

For our second set of experiments, we wanted to see if caching more embeddings meant better performance. For that, we set the batch size at 2048 and changed the GPU memory allocated for caching the hot embeddings. From Table \ref{table:2}, in each case when compared with the CPU-GPU baseline, we see major savings in the backward pass and optimization stages of one iteration, and the savings across different GPU memory settings are very similar. This is to be expected since our first experiment showed that caching hot embeddings significantly improves these stages thanks to decreased DRAM memory writes. However, forward time increases drastically and almost directly proportionally as the memory-cached hot embeddings increase. As a result, the forward time dominates the total iteration time. 

\begin{table}[htbp]
\caption{The absolute time in seconds for one iteration of SCARS with differing memory allocated for cached hot embeddings.}
\centering
\begin{tabular}{|c|c|c|c|c|}
\hline
Method & \multicolumn{4}{c|}{\textbf{One Iteration (s)}} \\
\cline{2-5} 
                               & Fwd & Bckwd & Optzn & Total Time \\ \hline
CPU-GPU baseline               & 20.37   & 19.14    & 54.17        & \textbf{93.68} \\
128MB                          & 26.65   & 4.13     & 0.39         & \textbf{31.23} \\
256MB                          & 50.88   & 5.36     & 0.41         & \textbf{56.65} \\
512MB                          & 108.46  & 5.79     & 0.42         & \textbf{114.75} \\
1024MB                         & 191.75  & 5.03     & 0.44         & \textbf{197.22} \\ \hline
\multicolumn{5}{l}{Note: Fwd, Bckwd, Optzn means Forwad, Backward, Optimization}

\end{tabular}
\label{table:2}
\end{table}

Such a pattern is caused by the increased size of the embedding lookup layer: As we cache more embeddings, the size of this layer increases. Since retrieving the embedding indices from a large lookup layer is slower, the forward time of the network, which is bounded by this lookup operation, increases. Hence, we conclude that we cannot just cache as many embeddings as we want and that we need a more intelligent way to cache hot embeddings. 

\begin{figure}[htbp]
\centerline{\includegraphics[width=0.5\textwidth]{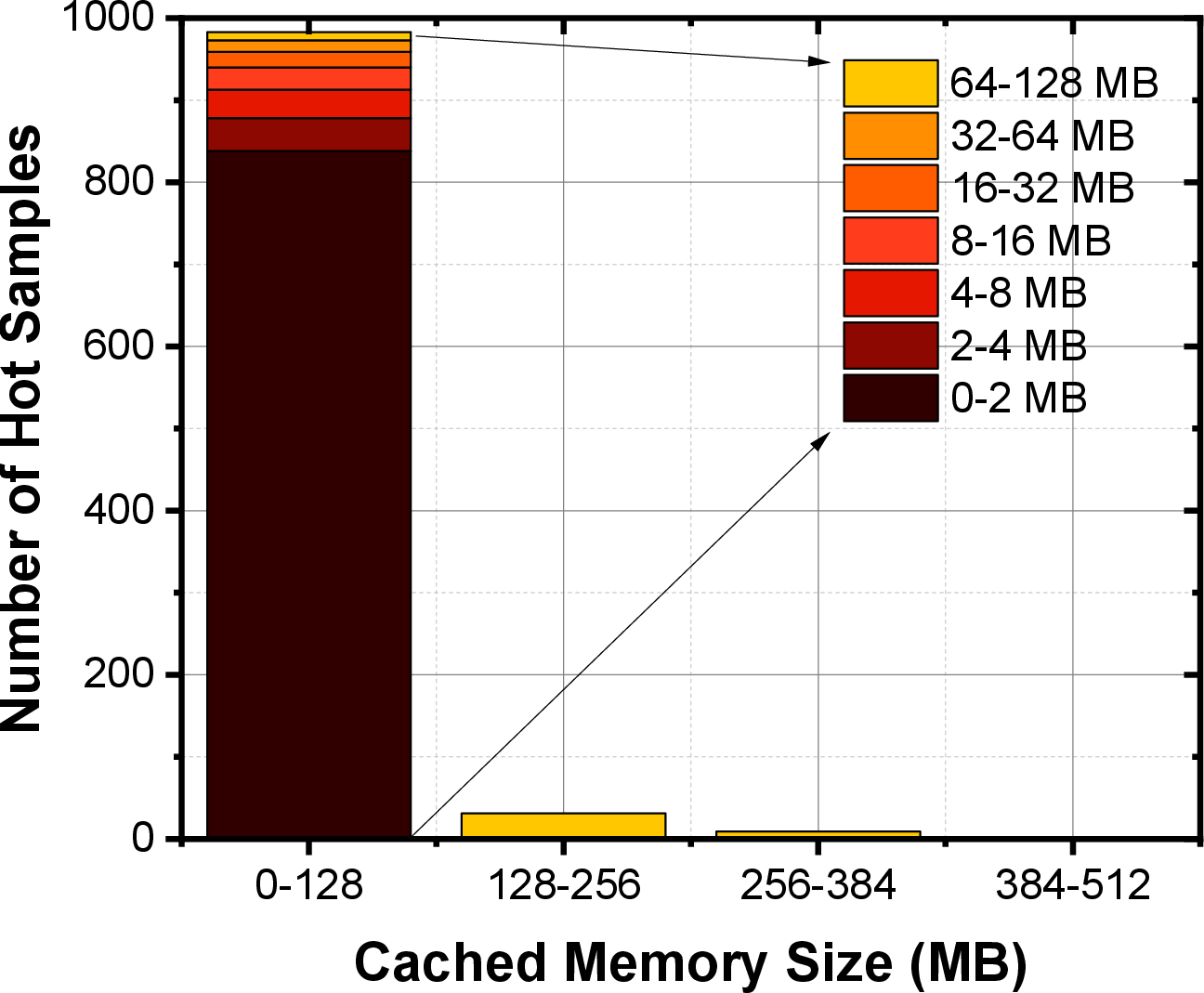}}
\caption{The number of samples using cached embeddings in a mini-batch of size 1024 from the Criteo Terabyte Dataset varies with the memory size. The 0-128MB range, containing the "hottest" embeddings, is used by almost all samples in a mini-batch, whereas the 128-256MB and 384-512MB ranges, representing progressively "colder" embeddings, see significantly less usage. The least hot 128MB range is seldom or never used. This data suggests that caching need not be indefinite, as caching colder embeddings eventually uses up lots of memory space without proportional usage benefits.}

\label{fig:hot_embedding_cache}
\end{figure}

Driven by the question of the optimal size of the cached hot embeddings, we analyzed how much of the cached embeddings in the 512MB setting from the previous experiment were frequently accessed by a batch of 1024 random training samples. We looked at the four 128MB portions of the 512MB, with the first portion containing the "hottest" cached embeddings and the last portion containing the least hot embeddings. As can be seen from Figure \ref{fig:hot_embedding_cache}, almost all of the samples in the batch accessed the first and the second portions while the third and the last portions were rarely or even never accessed. Thus, the last two 128MB portions are not really "hot" embeddings, and so caching them is unnecessary. All in all, increasing the memory for cached embeddings excessively causes us to cache cold embeddings and unnecessarily slow down the forward pass of the embedding layer. Thus, the best practice is to profile our data beforehand and see how much memory is really needed for cached embeddings. For our experiments, we decided to cache 256MB of hot embeddings.

Our next set of experiments with varied batch sizes showed that caching hot embeddings allows us to have larger batch sizes, as can be seen from Table \ref{table:3}. For a "normal" iteration, that is an iteration that accesses both the hot and the cold embeddings, a larger batch size means dramatically slower backward pass and optimization stages since a larger batch size means more DRAM writes which can be inferred from Table \ref{table:4}. Thus, state-of-the-art implementations use a batch size no larger than 2048. For a "hot" iteration, that is an iteration that accesses only the cached embeddings, we found that increased batch size doesn't affect the iteration time. The reason for such performance gain is that caching embeddings have major savings in backward and optimization stages already due to decreased DRAM writes and that since we now can limit the embeddings layer to an intelligent size (we choose 256MB for Criteo Terabyte Dataset), the time for forward pass doesn't change much with increased batch size.

\begin{table}[htbp]
\caption{The absolute time in seconds for one iteration of SCARS with differing batch sizes}
\centering
\begin{tabular}{|c|c|c|c|c|}
\hline
\textbf{Method} & \multicolumn{4}{c|}{\textbf{One Iteration (s)}} \\
\cline{2-5}
                            & Forward & Backward & Optimization & Total Time \\
\hline
2048                        & 50.88   & 5.36     & 0.41         & \textbf{56.65} \\
4096                        & 54.23   & 8.14     & 0.62         & \textbf{62.99} \\
8192                        & 51.60   & 10.43    & 0.88         & \textbf{62.91} \\
16384                       & 54.70   & 19.14    & 1.56         & \textbf{75.4}  \\
\hline
\end{tabular}
\label{table:3}
\end{table}

\begin{table}[htbp]
\caption{The absolute time in seconds for one iteration of the baseline CPU-GPU implementation with differing batch sizes}
\centering
\begin{tabular}{|c|c|c|c|c|}
\hline
\textbf{Method} & \multicolumn{4}{c|}{\textbf{One Iteration (s)}} \\
\cline{2-5}
                            & Fwd & Bckwd & Optzn & Total Time \\
\hline
2048                        & 20.37   & 19.14    & 54.17        & \textbf{93.68} \\
4096                        & 26.71   & 25.04    & 95.81        & \textbf{147.56} \\
8192                        & 40.06   & 74.81    & 149.52       & \textbf{264.39} \\
16384                       & 72.21   & 139.84   & 253.36       & \textbf{465.41} \\
\hline
\multicolumn{5}{l}{Note: Fwd, Bckwd, Optzn means Forwad, Backward, Optimization}

\end{tabular}
\label{table:4}
\end{table}

We also note the total time for one epoch of Criteo Terabyte with varying batch sizes in Table \ref{table:5} and see that caching hot embeddings allows us to use large batch sizes, which greatly improves the training time. However, after some very large batch sizes, increasing the batch size indefinitely doesn't correspond to equally greater performance gains. 

\begin{table}[htbp]
\caption{The absolute time in minutes for one epoch of baseline CPU-GPU implementation and SCARS with 256MB of cached hot embeddings}
\centering
\begin{tabular}{|c|ccccc|}
\hline
Method & \multicolumn{5}{c|}{Batch Size} \\
\cline{2-6}

                        & 2048   & 4096   & 8192   & 16384 & 32768 \\ \hline
CPU-GPU baseline        & 46.12 & 38.33 & 35.54 & 30.10 & -     \\ 
256MB Cached SCARS      & 27.12 & 15.10  & 7.54  & 4.52 & 4.31 \\ \hline
\end{tabular}
\label{table:5}
\end{table}

From Table \ref{table:6}, we can see that although the first few increases in the batch size reflect remarkable gains (a speed up of $~3.6\times$ from 2048 to 8192), the next few increases don't show an equally great gain (a speed up of $~1.7\times$ from 8192 to 16384). Hence, it is not essential to increase the batch size indefinitely further than some point.

\begin{table}[htbp]
\caption{Speedup ratio at p to q i.e. Total time in one iteration for batch size p to Total tie in one iteration for batch size q}
\centering
\begin{tabular}{|c|c|c|c|}
\hline
Batch Size & \multicolumn{3}{c|}{Speedup (s) for one iteration of SCARS} \\
\cline{2-4}
                            & 4096(p) & 8192(p)  & 16384(p) \\ \hline
2048(q)                     & 1.8     & 3.6       & 6.01     \\
4096(q)                     & -       & 2         & 3.342    \\
8192(q)                     & -       & -         & 1.7      \\
\hline
\multicolumn{4}{l}{}
\end{tabular}
\label{table:6}
\end{table}

Speedup(p/q) = (Absolute Time for one iteration at Batch Size p) /(Absolute Time for one iteration at Batch Size at q 

Next, we wanted to analyze how increasing the batch size of SCARS affects convergence and training accuracy. We keep the memory used for cached embeddings constant and this is considerably less than the device memory; therefore, we can technically indefinitely increase the batch size. Since our method mainly focuses on improving performance in terms of time and efficiency, we let models with different batch sizes train for an absolute time of 10 hours rather than using some certain epoch count since, as shown before, the time for an epoch for different batch sizes is not the same. These time frames, as a result, would mean more epochs for large batch sizes while it only corresponds to a few epochs for smaller ones. Our experiments showed that after you increase the batch size too much, although the speed and efficiency gains are remarkable, the model cannot converge as fast. Since the size of mini-batches essentially determines the frequency of updates, very large mini-batches correspond to fewer updates and therefore slower convergence; therefore, it is best to choose a large batch size reasonably since a slow convergence is suboptimal. This finding, when paired with our previous finding that increasing the batch size after some point doesn't correspond to equally great performance gains in terms of speed, leads us to such insights: Since increasing the batch size "too much" ceases to offer striking speed ups in training time and also slows down convergence, it is suboptimal to increase the batch size indefinitely. Besides, we found that too large batch size training is more prone to overfitting. 

Since after 10 hours, the models with reasonable batch sizes (2048-8192) are all close to convergence (they are very far into their training), their accuracy is all high and close to each other as well; hence, we can't really observe the speed at which the model initially gets close to a minimum, which usually happens in the first few iterations of the training. This initial stage is the defining factor for measuring the speed of convergence since the most accuracy gains happen here. Thus, we need to see which model has a faster initial stage than the others, and for that matter, we need a shorter time frame. We chose a time frame of 1 hour to repeat the experiment and observed if SCARS with large batch sizes converged faster. This time, however, the difference between the small and large batch sizes was greater and more stark. As can be seen from Table \ref{table:7}, we found that reasonably large batch size schemes converge faster than the model with a batch size of 2048; indeed, SCARS with a batch size of 8192 achieves $0.214\%$ increase in accuracy compared to 2048; similarly, 4096 achieves $0.06\%$ increase in accuracy. The model with a 2048 batch size can only close this accuracy gap after about 1.6 more hours of training. This further shows that SCARS is able to train faster with large batch sizes and is time-friendly. This could prove to be very beneficial for commercial scenarios where the training time is limited and for research purposes where you would want to get quick results (or any kind of situation in which waiting for one day is not optimal). 

\begin{table}[htbp]
\caption{Test accuracy after 1 and 10 hours of training}
\centering
\begin{tabular}{|c|c|c|c|c|c|c|}
\hline
   & \multicolumn{6}{c|}{Speedup (s) for one iteration of SCARS} \\
\cline{2-7}
                                   & 2048 & 4096 & 8192 & 16384 & 32768 & 65536 \\
\hline
1 hour                             & 80.510 & 80.573 & \textbf{80.724} & 80.457 & 80.066 & 79.587 \\
10 hours                           & 81.171 & 81.168 & \textbf{81.176} & 81.094 & 81.021 & 80.816 \\
\hline

\end{tabular}
\label{table:7}
\end{table}

Consequently, when the hot embeddings are intelligently cached and therefore when we can use a reasonably large batch size, our method greatly reduces the training time with a faster convergence by achieving up to ~$6\times$ speed up from Table \ref{table:6} compared to the CPU-GPU baseline implementation of DLRM. 

\section{Conclusion}
In this work, we develop a versatile framework that models the communication costs of various distributed systems. Leveraging this framework, we minimize the expected costs by applying effective communication strategies for large-scale recommendation system training. Specifically, our method utilizes overlapping embeddings to transmit only the unique embeddings along with their indices for each feature. Additionally, we leverage the GPU as a cache for frequently accessed embeddings. We explore several methods to determine which embeddings to store on the GPU and derive a theoretical solution addressing the tradeoff between batch size and the size of the cached embeddings, as well as theory-inspired experimental solutions. As a result, our method generalizes well across different datasets and systems with varying memory constraints. Furthermore, our PyTorch implementation demonstrates up to $6\times$ improvements in training and convergence times on large GPU systems using both the Criteo Terabyte and Alibaba User Behavior datasets.

\bibliographystyle{IEEEtran}
\bibliography{egbib}

\begin{thebibliography}{10}
\providecommand{\url}[1]{#1}
\csname url@samestyle\endcsname
\providecommand{\newblock}{\relax}
\providecommand{\bibinfo}[2]{#2}
\providecommand{\BIBentrySTDinterwordspacing}{\spaceskip=0pt\relax}
\providecommand{\BIBentryALTinterwordstretchfactor}{4}
\providecommand{\BIBentryALTinterwordspacing}{\spaceskip=\fontdimen2\font plus
\BIBentryALTinterwordstretchfactor\fontdimen3\font minus \fontdimen4\font\relax}
\providecommand{\BIBforeignlanguage}[2]{{%
\expandafter\ifx\csname l@#1\endcsname\relax
\typeout{** WARNING: IEEEtran.bst: No hyphenation pattern has been}%
\typeout{** loaded for the language `#1'. Using the pattern for}%
\typeout{** the default language instead.}%
\else
\language=\csname l@#1\endcsname
\fi
#2}}
\providecommand{\BIBdecl}{\relax}
\BIBdecl

\bibitem{fan2021dapple}
S.~Fan, Y.~Rong, C.~Meng, Z.~Cao, S.~Wang, Z.~Zheng, C.~Wu, G.~Long, J.~Yang, L.~Xia \emph{et~al.}, ``Dapple: A pipelined data parallel approach for training large models,'' in \emph{Proceedings of the 26th ACM SIGPLAN Symposium on Principles and Practice of Parallel Programming}, 2021, pp. 431--445.

\bibitem{jia2019beyond}
Z.~Jia, M.~Zaharia, and A.~Aiken, ``Beyond data and model parallelism for deep neural networks.'' \emph{Proceedings of Machine Learning and Systems}, vol.~1, pp. 1--13, 2019.

\bibitem{zhao2023pytorch}
Y.~Zhao, A.~Gu, R.~Varma, L.~Luo, C.-C. Huang, M.~Xu, L.~Wright, H.~Shojanazeri, M.~Ott, S.~Shleifer \emph{et~al.}, ``Pytorch fsdp: experiences on scaling fully sharded data parallel,'' \emph{arXiv preprint arXiv:2304.11277}, 2023.

\bibitem{gholami2022survey}
A.~Gholami, S.~Kim, Z.~Dong, Z.~Yao, M.~W. Mahoney, and K.~Keutzer, ``A survey of quantization methods for efficient neural network inference,'' in \emph{Low-Power Computer Vision}.\hskip 1em plus 0.5em minus 0.4em\relax Chapman and Hall/CRC, 2022, pp. 291--326.

\bibitem{dong2019hawq}
Z.~Dong, Z.~Yao, A.~Gholami, M.~W. Mahoney, and K.~Keutzer, ``Hawq: Hessian aware quantization of neural networks with mixed-precision,'' in \emph{Proceedings of the IEEE/CVF international conference on computer vision}, 2019, pp. 293--302.

\bibitem{liu2023noisyquant}
Y.~Liu, H.~Yang, Z.~Dong, K.~Keutzer, L.~Du, and S.~Zhang, ``Noisyquant: Noisy bias-enhanced post-training activation quantization for vision transformers,'' in \emph{Proceedings of the IEEE/CVF Conference on Computer Vision and Pattern Recognition}, 2023, pp. 20\,321--20\,330.

\bibitem{cheng2016wide}
H.-T. Cheng, L.~Koc, J.~Harmsen, T.~Shaked, T.~Chandra, H.~Aradhye, G.~Anderson, G.~Corrado, W.~Chai, M.~Ispir \emph{et~al.}, ``Wide \& deep learning for recommender systems,'' in \emph{Proceedings of the 1st workshop on deep learning for recommender systems}, 2016, pp. 7--10.

\bibitem{zhao2022analysis}
L.~Zhao, Z.~Dong, and K.~Keutzer, ``Analysis of quantization on mlp-based vision models,'' \emph{arXiv preprint arXiv:2209.06383}, 2022.

\bibitem{naumov2019deep}
M.~Naumov, D.~Mudigere, H.-J.~M. Shi, J.~Huang, N.~Sundaraman, J.~Park, X.~Wang, U.~Gupta, C.-J. Wu, A.~G. Azzolini \emph{et~al.}, ``Deep learning recommendation model for personalization and recommendation systems,'' \emph{arXiv preprint arXiv:1906.00091}, 2019.

\bibitem{guan2019post}
H.~Guan, A.~Malevich, J.~Yang, J.~Park, and H.~Yuen, ``Post-training 4-bit quantization on embedding tables,'' \emph{arXiv preprint arXiv:1911.02079}, 2019.

\bibitem{zhou2024dqrm}
Y.~Zhou, Z.~Dong, E.~Chan, D.~Kalamkar, D.~Marculescu, and K.~Keutzer, ``Dqrm: Deep quantized recommendation models,'' \emph{arXiv preprint arXiv:2410.20046}, 2024.

\bibitem{yin2021tt}
C.~Yin, B.~Acun, C.-J. Wu, and X.~Liu, ``Tt-rec: Tensor train compression for deep learning recommendation models,'' \emph{Proceedings of Machine Learning and Systems}, vol.~3, pp. 448--462, 2021.

\bibitem{li2024embedding}
S.~Li, H.~Guo, X.~Tang, R.~Tang, L.~Hou, R.~Li, and R.~Zhang, ``Embedding compression in recommender systems: A survey,'' \emph{ACM Computing Surveys}, vol.~56, no.~5, pp. 1--21, 2024.

\bibitem{mangalam2022reversible}
K.~Mangalam, H.~Fan, Y.~Li, C.-Y. Wu, B.~Xiong, C.~Feichtenhofer, and J.~Malik, ``Reversible vision transformers,'' in \emph{Proceedings of the IEEE/CVF Conference on Computer Vision and Pattern Recognition}, 2022, pp. 10\,830--10\,840.

\bibitem{jain2020checkmate}
P.~Jain, A.~Jain, A.~Nrusimha, A.~Gholami, P.~Abbeel, J.~E. Gonzalez, I.~Stoica, and K.~Keutzer, ``Checkmate: Breaking the memory wall with optimal tensor rematerialization,'' in \emph{Proceedings of Machine Learning and Systems}, vol.~2, 2020, pp. 497--511.

\bibitem{shang2023pb}
Y.~Shang, Z.~Yuan, Q.~Wu, and Z.~Dong, ``Pb-llm: Partially binarized large language models,'' \emph{arXiv preprint arXiv:2310.00034}, 2023.

\bibitem{li2023qft}
Z.~Li, X.~Liu, B.~Zhu, Z.~Dong, Q.~Gu, and K.~Keutzer, ``Qft: Quantized full-parameter tuning of llms with affordable resources,'' \emph{arXiv preprint arXiv:2310.07147}, 2023.

\bibitem{shoeybi2019megatron}
M.~Shoeybi, M.~Patwary, R.~Puri, P.~LeGresley, J.~Casper, and B.~Catanzaro, ``Megatron-lm: Training multi-billion parameter language models using model parallelism,'' \emph{arXiv preprint arXiv:1909.08053}, 2019.

\bibitem{agarwal2023bagpipe}
S.~Agarwal, C.~Yan, Z.~Zhang, and S.~Venkataraman, ``Bagpipe: Accelerating deep recommendation model training,'' in \emph{Proceedings of the 29th Symposium on Operating Systems Principles}, 2023, pp. 348--363.

\bibitem{zhao2020distributed}
W.~Zhao, D.~Xie, R.~Jia, Y.~Qian, R.~Ding, M.~Sun, and P.~Li, ``Distributed hierarchical gpu parameter server for massive scale deep learning ads systems,'' \emph{Proceedings of Machine Learning and Systems}, vol.~2, pp. 412--428, 2020.

\bibitem{ishkhanov2020time}
T.~Ishkhanov, M.~Naumov, X.~Chen, Y.~Zhu, Y.~Zhong, A.~G. Azzolini, C.~Sun, F.~Jiang, A.~Malevich, and L.~Xiong, ``Time-based sequence model for personalization and recommendation systems,'' \emph{arXiv preprint arXiv:2008.11922}, 2020.

\end{thebibliography}

\end{document}